\documentstyle[12pt]{article}

\begin{document}

\begin{titlepage}
\rightline{CERN-TH/95-56}
\rightline{RU-95-11}
\bigskip
\bigskip
\begin{center}
{\LARGE\bf{An Anti-WKB Approximation}}

\bigskip
\bigskip
\bigskip

{\large J. B. Bronzan}

\bigskip
CERN, CH-1211 Geneva 23, Switzerland\\
\smallskip
and\\
\smallskip
Department of Physics and Astronomy\\
Rutgers University\\
Piscataway, NJ 08855-0849, USA\\
email: bronzan@physics.rutgers.edu

\end{center}

\bigskip

\abstract{
\noindent
In the WKB approximation the $\nabla^2 S$ term in Schr\"odinger's equation is
subordinate to the $|{\bf\nabla}S|^2$ term.  Here we study an {\it anti-WKB
approximation} in which the $\nabla^2 S$ term dominates (after a guess for
$S$ is supplied).  Our approximation produces only the nodeless ground state
wavefunction, but it can be used in potential problems where the potential
is not symmetric, and in problems where there are many degrees of freedom.  As
a test, we apply the method to potential problems, including the hydrogen and
helium atoms, and to $\phi^4$ field theory.}

\vfill

\leftline{CERN-TH/95-56}
\leftline{RU-95-11}
\leftline{March 1995}

\end{titlepage}

\section{Introduction}
\label{Introduction}
\setcounter{equation}{0}

     In quantum mechanics the wavefunction is sometimes written in the form
\begin{equation}
\psi ({\bf r})=\exp S({\bf r}).
\end{equation}
Schr\"odinger's equation then takes the form
\begin{equation}
\nabla^2 S+|{\bf\nabla}S|^2={{2m}\over{\hbar^2}}\left [ V({\bf r})-E\right ].
\label{eq:schrod1}
\end{equation}
The WKB approximation ensues when the term $\nabla^2 S$ is dropped in leading
order and later incorporated as a correction.\cite{landau1}  It is commonly
understood that this step is justified in the semi-classical regime where the
gradient of the de Broglie wavelength, ${\bf\nabla}\lambda$, has a magnitude
much smaller than one, and the wave function oscillates many times over
distances that characterize the variation of the potential.\par
     In this paper we consider the alternate ordering of terms where the
term $|{\bf\nabla}S|^2$ is dropped in leading order and later treated as a
correction.  This is an {\it anti-WKB approximation}.  As we shall see, the
anti-WKB approximation is valid deep in the quantum regime where the
wavefunction is nodeless and therefore changes over a length that is at least
as great as that which characterizes the variation of the potential.  For
a particle moving in a potential, such a wavefunction describes a bound
state.  In a bosonic field theory it is describes the ground state.

     The anti-WKB approximation is not as straightforward as the WKB
approximation because it is not generally true that $|\nabla^2 S|>>
|{\bf\nabla} S|^2$ for bound state wavefunctions.  To see this explicitly,
assume the potential vanishes when $r>a$.  Then for $r>a$,
\begin{equation}
S=S_0-Kr-{{D-1}\over 2}\ln (r/r_0); \qquad K\equiv\sqrt{-{{2mE}\over
{\hbar^2}}},
\label{eq:asymptotics}
\end{equation}
where motion is in $D$ spatial dimensions.  Comparing the gradient and
Laplacian applied to this expression, we find that the anti-WKB approximation
is justified only when $D=1$ and $K=0$ (or at least $K\sim 0$.)  This is
the case of a weakly bound particle moving in one dimension.  (Recall that in
one dimension there is always one bound state in an attractive potential, no
matter how weak the potential.)  This version of the anti-WKB approximation
has been developed.\cite{bronzan}  One obtains an expression for the
energy of the bound state as a series of integrals over powers of the
potential.  The relative size of the $n$-th term is proportional to the
$n$-th power of the parameter ${V_0ma^2}/\hbar^2$, where $V_0$ is the strength
of the potential and $a$ its range.  We see here an expected contrast with
the WKB approximation: a series of {\it decreasing} powers of Plack's
constant.  Despite this, if the potential is such as to make the parameter
small, the approximation succeeds.

     The straightforward case just described is quite different from the
anti-WKB approximation considered in this paper, where we treat the general
case $D\ne 1$, $K\ne 0$.  We evade the ``no-go'' conclusion, above, by
supplying an initial guess for $S$.  Thus
we write
\begin{equation}
S({\bf r})=F({\bf r})+T({\bf r}),
\end{equation}
where the variational seed $F$ has the asymptotic form \ref{eq:asymptotics},
and is chosen to be an initial guess for $S$.  The correction $T$ is now
determined by Schr\"odinger's equation:
\begin{eqnarray}
\nabla^2 T_n+2{\bf\nabla}F{\bf\cdot\nabla}T_n={{2m}\over{\hbar^2}}\left [
V({\bf r})-E_n\right ]-\nabla^2 F-|{\bf\nabla}F|^2-|{\bf\nabla} T_{n-1}|^2;
\nonumber\\ T_{-1}=0.
\label{eq:schrod2}
\end{eqnarray}
The index $n=0,1\dots$ labels the successive approximations to $T$ and hence
$S$.  The initial approximation ignores the gradient of $T$; the next
approximation uses the initial approximation for the gradient, and so forth.
Of course, ignoring the gradient of $T$ is different than ignoring the
gradient of S, much of which appears in Eq.~\ref{eq:schrod2} in the term
$|{\bf\nabla}F|^2$; nevertheless, we still call this the anti-WKB
approximation.  Note that for each $n$ we obtain a different
approximation for the bound state energy, $E_n$.

     In Section 2 we demonstrate that this sequence of approximations is
formally convergent provided $F$ is well chosen.  It is useful to state what
we find.  Assume that our variational seed differs from $S$ by a function
scaled by a small pararmeter $\epsilon$:
\begin{equation}
F({\bf r})=S({\bf r})-\epsilon S_1({\bf r}).
\end{equation}
It follows that the exact $T$ is $\epsilon S_1$.  In Section 2 we show that
\begin{equation}
E_n=E+\epsilon^{n+2}A_n;\qquad T_n=\epsilon S_1+\epsilon^{n+2}R_n.
\label{eq:epsilon}
\end{equation}
Here $E$ is the true bound state energy, and the factors $A_n$ and $R_n$ are
finite at $\epsilon =0$.  The convergence is formal because we have not given
estimates for the $n$-dependence of $A_n$ and $R_n$.  If these factors
increase with n faster than an exponential, the sequence does not not
converge, but is instead asymptotic.

     The convergence estimates of Eq.~\ref{eq:epsilon} depend on the fact
that the term $T_{n-1}$ appears quadratically on the right hand side of
Eq.~\ref{eq:schrod2}.  When $F$ differs little from $S$, treating the
gradient as a correction is justified.  We again emphasize the difference
between what is done in Ref.~\cite{bronzan} and what we do here.  This is
not a weak potential approximation, nor is it an expansion in inverse powers
of $\hbar$.  We rely on the availability of a reasonable guess for the
variational seed $F$.  The guess must also be such
that the integrals and Green functions we shall encounter can be computed.

     The wavefunction arising from our construction is nodeless.  The reason
is that if $\psi$ vanishes on the surface $f({\bf r})=0$, there will be a
term $\ln f({\bf r})$ in $S$.  In the neighborhood of the node, this
logarithm dominates $S$, and we must explicitly incorporate it into $F$ if we
are obtain a wave function with a node.  Note that we must specify the
surface $f=0$, which is known only under special circumstances.  We do not
consider such cases here; our $F$'s will be smooth, and we therefore limit
ourselves to nodeless wavefunctions.  This means that the energies $E_n$ are
approximations to the ground state energy $E$.

     Like the WKB approximation, the anti-WKB approximation is
nonperturbative; it does not require the presence of a small parameter in
the Hamiltonian.  Its major limitation is that it is restricted to the ground
state wave function.  But the anti-WKB approximation has this important
advantage: Equation \ref{eq:schrod2} for $T_n$ is linear.  We will see that
it can be solved readily in many cases of interest, including particles
moving in three dimensions in asymmetric potentials, and many-body problems
like bosonic lattice field theory.  In the latter problem the method can to
extended
to study some of the vacuum state correlation functions that are of central
importance in field theory.  All these possibilites are closed to the WKB
approximation, which is generally unmanageable except for a particle whose
motion effectively reduces to one dimension.  It should be noted that the
anti-WKB approximation is less useful for problems involving identical
fermions because the Pauli principle makes the nodeless state unphysical for
systems of more than two spin 1/2 particles.

     In Section 2 we present a theoretical development of the anti-WKB
approximation.  We include the solution of the dynamical
equation~\ref{eq:schrod2}, the convergence of the sequence of
approximations, and several other matters.

     In following sections the anti-WKB approximation is applied to a
number of problems to show that it works in increasingly complex
situations.  In Section 3 we study a spherically symmetric square well whose
ground state is known by elementary methods.  We find that $E_0$ gives only
62\% of the correct binding energy, but the next approximation, $E_1$, gives
96\%.  We develop Green functions required for the application of
Eq.~\ref{eq:schrod2} to potential problems where the potential is
nonspherical.  In Section 4 we study long range potentials, with the hydrogen
atom as a particularly simple subcase.  We finish by applying the
anti-WKB approximation to many-body problems: $\phi^4$ field theory in
Section 5, and the helium atom in Section 6.  Conclusions are presented in
Section 7.

\section{Anti-WKB Equations}
\setcounter{equation}{0}

     The energies in Eq.~\ref{eq:schrod2} are determined by a general
requirement.  Consider the surface integral
\begin{equation}
\oint_S d{\bf A\cdot}\left [e^{2F}{\bf\nabla}T_n\right ]=\int_V dV{\bf\nabla
\cdot}\left [e^{2F}{\bf\nabla}T_n\right ].
\end{equation}
As $S$ is expanded to infinity, the surface integral decreases to zero
because of the exponential fall of the factor $e^{2F}$.  Using Eq.~\ref
{eq:schrod2} we obtain the eigenvalue equation determining $E_n$:
\begin{equation}
0=\int dV e^{2F}\left \{ {{2m}\over{\hbar^2}}\left [ V({\bf r})-E_n\right ]-
\nabla^2 F-|{\bf\nabla}F|^2-|{\bf\nabla} T_{n-1}|^2\right \}.
\label{eq:eigenvalue}
\end{equation}

     We can now assemble the equations to derive the convergence results
\ref{eq:epsilon}.  The true
ground state energy $E$ is determined by Eq.~\ref{eq:eigenvalue} with the
replacements $E_n\rightarrow E$, $T_{n-1}\rightarrow T$.  Subtracting
equations,
\begin{eqnarray}
0=\int dV e^{2F}\left [{{2m}\over {\hbar^2}}\left (E_n-E\right )+
|{\bf\nabla} T_{n-1}|^2-|{\bf\nabla} T|^2\right ],\label{eq:eigenvalue2} \\
E_n=E+{\hbar^2\over{2m}}{{\int dV e^{2F}\left [|{\bf\nabla}T|^2
-|{\bf\nabla}T_{n-1}|^2\right ]}\over{\int dV e^{2F}}}.\nonumber
\end{eqnarray}
We now express the $n$-th order factors in terms of their limiting values and
a deviation:
\begin{eqnarray}
E_n=E+a_n,\\
T_n=-\epsilon S_1+r_n.\nonumber
\end{eqnarray}
Then Eq.~\ref{eq:eigenvalue2} becomes
\begin{eqnarray}
a_0=&&\epsilon^2{\hbar^2\over{2m}}{{\int dV e^{2F}|{\bf\nabla} S_1|^2}\over
{\int dV e^{2F}}},\label{eq:as} \\
a_n=&&{\hbar^2\over{2m}}{{\int dV e^{2F}\left [ 2\epsilon {\bf\nabla}S_1{\bf
\cdot\nabla}r_{n-1}-|{\bf\nabla}r_{n-1}|^2\right ]}\over{\int dV e^{2F}}},
\qquad (n\ge 1).\nonumber
\end{eqnarray}
We next use Eq.~\ref{eq:schrod2} and the analogous equation for $T=\epsilon
S_1$.  Subtracting the equations
\begin{eqnarray}
\nabla^2r_0+2{\bf\nabla}F{\bf\cdot\nabla}r_0=&&-{{2m}\over\hbar^2}a_0+
\epsilon^2|{\bf\nabla}S_1|^2,\label{eq:rs} \\
\nabla^2r_n+2{\bf\nabla}F{\bf\cdot\nabla}r_n=&&-{{2m}\over\hbar^2}a_n+
2\epsilon{\bf\nabla}S_1{\bf\cdot\nabla}r_{n-1}-|{\bf\nabla}r_{n-1}|^2.
\qquad (n\ge 1).\nonumber
\end{eqnarray}
It follows that Eq.~\ref{eq:epsilon} holds for $n=0$.  Furthermore,
if \ref{eq:epsilon} holds for $n$, then from Eqs.~\ref{eq:as}, \ref{eq:rs} we
see that it holds for $n+1$, and the result is established by induction.

     Energy $E_n$ requires $T_{n-1}$ for its computation.  This information
can be used to find a better estimate of the state energy from the expectation
\begin{equation}
{\cal E}_n={{<\psi|H|\psi >}\over{<\psi|\psi>}},\qquad \psi=\exp (F+T_{n-1}).
\end{equation}
The error in $\psi$ is $O(\epsilon^{n+1})$; variational argument~\cite{
LL2} gives
\begin{equation}
{\cal E}_n-E=O(\epsilon^{2n+2});\qquad {{{\cal E}_n-E}\over{|E_n-E|}}=O(
\epsilon^n).
\end{equation}

     Eq.~\ref{eq:schrod2} may be solved using the Green function satisfying
\begin{eqnarray}
LG({\bf r},{\bf r'})=\delta ({\bf r}-{\bf r'}),\label{eq:Greenfun} \\
L=\nabla^2+2({\bf\nabla}F){\bf\cdot\nabla}.\nonumber
\end{eqnarray}
Then
\begin{equation}
T_n({\bf r})=\int dV' G({\bf r},{\bf r'})\left \{ {{2m}\over\hbar^2}\left [
V-E_n\right ]-\nabla^2 F-|{\bf\nabla}F|^2-|{\bf\nabla}T_{n-1}|^2\right \}_{\bf
r'}.
\label{eq:tee}
\end{equation}

     It is illuminating to expand the Green function in terms of the
orthonormal
eigenfunctions of L, because the argument leads to the eigenvalue equation
\ref{eq:eigenvalue} in a different way.
\begin{equation}
L\phi_k=\mu_k\phi_k;\qquad \int dV e^{2F}\phi_{k1}\phi_{k2}=\delta_{k1,k2}.
\end{equation}
Then
\begin{equation}
G({\bf r},{\bf r'})=\sum_k{1\over{\mu_k}}\phi_k({\bf r})\phi_k({\bf r'})
e^{2F({\bf r'})}.
\label{eq:Greenexp}
\end{equation}
But there is a problem: $L$ annihilates a constant function, so there is a
normalized eigenfunction of $L$ with eigenvalue zero:
\begin{equation}
\phi_0({\bf r})=N,\quad N=\left [\int dV e^{2F}\right ]^{-1/2};\quad \mu_0
=0.
\end{equation}
The contribution of this zero mode to G is infinite.  Nonetheless, $T_n$ in
Eq.~\ref{eq:tee} is finite if the projection of the driving term onto the
zero mode vanishes.  The condition for that is the eigenvalue equation
\ref{eq:eigenvalue}.

     The eigenvalue equation for $E_1$ can be simplified.  We need the
integral
\begin{equation}
-\int dV e^{2F}|T_0|^2=-\int dV{\bf\nabla\cdot}\left [e^{2F}T_0{\bf\nabla}T_0
\right ]+\int dVe^{2f}T_0{\bf\nabla\cdot}\left [e^{2F}{\bf\nabla}T_0\right ]
\end{equation}
The first integral on the right vanishes by the divergence theorem, and the
second may be transformed using Eq.~\ref{eq:schrod2}.
\begin{equation}
-\int dV e^{2F}|T_0|^2=\int dV e^{2F}T_0D=\int dVdV'e^{2F({\bf r})}D({\bf r})
D({\bf r'})G({\bf r},{\bf r'}),
\end{equation}
where
\begin{equation}
D\equiv {{2m}\over\hbar^2}\left [V-E_0\right ]-\nabla^2 F-|{\bf\nabla}F|^2.
\end{equation}
The quantization condition for $E_1$ reads
\begin{eqnarray}
0=&&\int dVe^{2F}\left [D+{{2m}\over\hbar^2}(E_0-E_1)\right ]+\int dVdV'e^
{2F({\bf r})}D({\bf r})D({\bf r'})G({\bf r},{\bf r'});\nonumber \\
E_1=&&E_0+{\hbar^2\over{2m}}{{\int dVdV'e^{2F({\bf r})}D({\bf r})D({\bf r'})
G({\bf r},{\bf r'})}\over{\int dVe^{2F}}}.
\label{eq:E1equat}
\end{eqnarray}

     When $V({\bf r})$ depends on a single coordinate, all differential
equations involved in constructing the Green function can be solved.  Consider
the case where $V$, $F$ and $T$ depend only on the radial coordinate, as
happens in Section 3.  Then the Green function satisfies the ordinary
differential equation
\begin{eqnarray}
Lg(r,r')={1\over r'^2}\delta (r-r'),\\
L={d^2\over{dr^2}}+2\left( {1\over r}+{{dF}\over{dr}}\right ){d\over{dr}}.
\nonumber
\end{eqnarray}
Rather than expand in eigenfunctions, we construct $g$ from the two solutions
of $LR=0$, which are known.
\begin{equation}
R^+(r)=1;\qquad R^-(r)=\int^r_{r_0}{{dr'e^{-2F(r')}}\over{r'^2}}.
\label{eqn:explrad}
\end{equation}
The Green function is
\begin{equation}
g(r,r')=R^+(r_>)R^-(r_<)/r'^2W(r')=-R^+(r_>)R^-(r_<)e^{2F(r')},
\end{equation}
where W is the Wronskian
\begin{equation}
W(R^+,R^-)\equiv {{dR^+}\over{dr}}R^--R^+{{dR^-}\over{dr}}=
-{{e^{-2F(r)}}\over{r^2}}.
\end{equation}
Now
\begin{eqnarray}
T_n(r)=\int_0^{\infty}dr'r'^2g(r,r')\Biggl \{{{2m}\over\hbar^2}\left [V-E_n
\right ]\\
-F''-{2\over{r'}}F'-(F')^2-(T'_{n-1})^2\Biggr \}_{r'}.\nonumber
\end{eqnarray}
The eigenvalue equation emerges here, not through a zero mode, but through
the bad asymptotic behavior of $R^-$.  Near $r=0$, $R^-(r)\sim e^{-2F(0)}/r$,
and
\begin{eqnarray}
T_n(r)\sim {{e^{-2F(0)}}\over{r}}\int_0^{\infty}dr'r'^2e^{2F(r')}\Biggl \{
{{2m}\over\hbar^2}\left [V-E_n\right ]\label{eqn:sphereigen} \\
-F''-{2\over{r'}}F'-(F')^2-(T'_{n-1})^2\Biggr \}_{r'}.\nonumber
\end{eqnarray}
To keep $T_n$ finite at $r=0$, the eigenvalue integral must vanish; when it
does, $T_n(0)=0$.

\section{Spherical Square Well Potential}
\setcounter{equation}{0}

     We begin our study of the anti-WKB approximation by applying it to the
case of the spherically symmetric potential
\begin{eqnarray}
V(r)=\cases{-V_0&$(r<a)$;\cr
               0&$(r>a)$.\cr}
\end{eqnarray}
The radial Schr\"odinger's equation for s-wave bound states can be solved, and
the energies are determined by the equation\cite{landau2}
\begin{equation}
ka\cot ka=-Ka;\qquad ka=\sqrt{{{2ma^2}\over{\hbar^2}}V_0-(Ka)^2};\qquad
E=-{{\hbar^2K^2}\over{2m}}.
\label{eq:exactenergy}
\end{equation}
There is no bound state unless $V_0$ exceeds the threshold strength
\begin{equation}
[V_0]_{thresh}={{\hbar^2}\over{2ma^2}}\left ({{\pi}\over 2}\right )^2C_1;
\qquad C_1=1.0,
\end{equation}
and for $V_0$ just above threshold strength, the energy of the ground state is
\begin{equation}
E={{ma^2}\over{2\hbar^2}}(V_0-[V_0]_{thresh})^2C_2;\qquad C_2=1.0.
\end{equation}

     To compute $E_0$ we use the eigenvalue integral \ref{eqn:sphereigen}.
To get some notion of the dependence of $E_0$ on the variational seed, we
try three of them, each of which depends on a single parameter $r_0$:
\begin{eqnarray}
F_1(r)=-Kr-\ln (r/r_0+1);\nonumber\\
F_2(r)=-Kr-\ln\sqrt{(r/r_0)^2+1};\label{eqn:threeFs} \\
F_3(r)=\cases{1-(K+{1\over{r_0}})r,&$(r<r_0);$\cr
               -Kr-\ln (r/r_0),&$(r>r_0)$.\cr}\nonumber
\end{eqnarray}
($F_3$ is chosen so that the function and its derivative are continuous at
$r=r_0$.)  Using these variational seeds, we obtain the results summarized
in Table 2.1.
\vfill
\eject
\hrule
\begin{tabbing}
Thisist\=hewaytosettabs;\=topreventprinting\=,endthelinewitha\=killstateme\kill
       \>$F(r)$         \>$r_0/a$          \>$C_1$           \>$C_2$       \\
       \>               \>                 \>                \>            \\
       \>               \>                 \>                \>            \\
       \>$S(r)$ (exact) \>                 \> 1.0            \> 1.0        \\
       \>               \>                 \>                \>            \\
       \>$F_1(r)$       \>0.638            \>1.14           \>0.127       \\
       \>               \>                 \>                \>            \\
       \>$F_2(r)$       \>0.923            \>1.09           \>1.37        \\
       \>               \>                 \>                \>            \\
       \>$F_3(r)$       \>1.38             \>1.03           \>0.826
\end{tabbing}
\hrule
\vspace{.1in}
{\it Table 2.1.  The optimum values of $r_0/a$ and the coefficients $C_1$
and $C_2$ for the three \vspace{.2in}choices for $F(r)$.}\hfill\break
Note that $E_0={\cal E}_0$ is a variational energy; in each case
the parameter
$r_0/a$ has been chosen to minimize $E_0$.  We see that the results are
sensitive to $F$.  $F_3$ gives a threshold potential strength that is about
3\% high, and it also gives the best estimate for the bound state energy near
threshold.  We use $F_3$ in the rest of Section 3.

     To explore further we choose a potential strength well above threshold:
$2ma^2V_0/\hbar^2=3.0$  The bound state energy is given by Eq.~\ref
{eq:exactenergy} to be $E=-0.0613\hbar^2/2ma^2$.  We find that at this
potential strength $E_0$ is minimized for $r_0/a=1.505$ (in the threshold
calculation the value was 1.37).  We find $E_0=-0.0379\hbar^2/2ma^2.$  A
measure of the (mediocre) quality of this result is the ratio $E_0/E=0.619$.

     We next compute $E_1$.  For this one degree-of-freedom problem, we
use Eq.~\ref{eq:eigenvalue} because a simple formula for $dT_n/dr$ is
available:
\begin{eqnarray}
{{dT_n}\over{dr}}={{e^{-2F(r)}}\over{r^2}}\int_0^r dxx^2e^{2F(x)}\Biggl \{
{{2m}\over\hbar^2}[V(x)-E_n]\nonumber \\
-F''(x)-{2\over x}F'(x)-[F'(x)]^2-[T'_{n-1}(x)]^2\Biggr \}.
\end{eqnarray}
We now obtain much improved results: $E_1=-0.0589\hbar^2/2ma^2$, and
$E_1/E=0.960$.  Almost 90\% of the error in $E_0$ has been removed in
$E_1$.

     One of the properties of the anti-WKB approximation we have
emphasized is that it can be applied to potential problems where $V({\bf r})$
has a general dependence on ${\bf r}$.  This statement is qualified by
the requirement that $F({\bf r})$ be such that Eq.~\ref{eq:Greenfun} can be
solved for $G$.  One case where this can be done is when $V$ is {\it nearly}
spherically symmetric. Then we choose $F$ to depend on $r$, relying on the
$T_n$ to supply the nonspherical corrections.  Note that the anti-WKB method
is more flexibile than the method of separation of coordinates in that it is
not necessary for surfaces of constant potential to exactly fall on surfaces
of constant coordinate.  Approximate coincidence suffices.
Below we use separation of coordinates to construct $G$ in spherical
coordinates.

     When $F$ depends only on $r$, we write $G$ in the form
\begin{equation}
G({\bf r},{\bf r'})=\sum_{\ell =0}^\infty g_\ell (r,r')\sum_{m=-\ell}^{m=\ell}
Y_{\ell m}(\theta ,\phi )Y^*_{\ell m}(\theta ' ,\phi '),
\end{equation}
where the radial Green function satisfies
\begin{equation}
{{\partial^2 g_{\ell}(r,r')}\over{\partial r^2}}+2\left ({1\over r}+{{dF}\over
{dr}}\right ){{\partial g_{\ell}(r,r')}\over{\partial r}}-{{\ell (\ell +1)}.
\over r^2}g_{\ell}(r,r')={1\over r^2}\delta (r-r').
\end{equation}
This Green function is
\begin{equation}
g_{\ell}(r,r')=R^+_{\ell}(r_>)R^-_{\ell}(r_<)e^{2F(r')}/C_\ell,
\end{equation}
where $R^{+,-}$ satisfy the homogeneous equation
\begin{equation}
{{d^2R_{\ell}}\over{dr^2}}+2\left ({1\over r}+{{dF}\over{dr}}\right ){{dR_\ell}
\over{dr}}-{{\ell (\ell+1)}\over r^2}R_{\ell}=0.
\label{eqn:ODE}
\end{equation}
The factor $C_\ell$ comes from the Wronskian $W(R^+_{\ell},R^-_{\ell})=
C_{\ell}e^{-2F(r)}/r^2$.

     For $\ell =0$, Eq.~\ref{eqn:ODE} has the same solutions we found for the
case of spherical symmetry, and we can choose $R^{+,-}_0=R^{+,-}$ of Eq.~\ref
{eqn:explrad}.  We again encounter the eigenvalue integral
\ref{eqn:sphereigen} with this change: Only the $s$-wave projection of $V$
enters into the determination of $E_0$.  This is in keeping with our
assumption that $V$ is nearly spherically symmetric.  The higher $E_n$'s
receive contributions from all of the angular projections of $V$.

     The solutions of Eq.~\ref{eqn:ODE} for $\ell >0$ require further
analysis.  We first note that near $r=0$ they behave like $r^\ell$ or
$r^{-\ell -1}$, and we must choose $R^-_{\ell}$ to have the nonsingular
behavior $r^{\ell}$.  At large $r$, solutions of Eq.~\ref{eqn:ODE} behave
like those of Eq.~\ref{eqn:explrad} because the last term in Eq.~\ref{eqn:ODE}
is unimportant at large $r$.  We must choose the solution behaving
like a constant so that $R^{+}$ remains finite at large $r$.

     In the case of $F_3(r)$, Eq.~\ref{eqn:ODE} has solutions that are
familiar functions.  In the interval $r<r_0$, these are confluent
hypergeometic functions:
\begin{eqnarray}
R_{\ell 1}=r^{\ell}\Phi (\ell, 2\ell +2,2Kr+2r/r_0), \\
R_{\ell 2}=r^{-\ell -1}\Phi (-\ell -1, -2\ell, 2Kr+2r/r_0).\nonumber
\end{eqnarray}
In the interval $r>r_0$ they are modified spherical Bessel functions:
\begin{equation}
R_{\ell 3}=\sqrt{r}e^{Kr}I_{\ell +1/2}(Kr),\qquad R_{\ell 4}=\sqrt{r}e^{Kr}
K_{\ell +1/2}(Kr).
\end{equation}
We then have the solutions
\begin{eqnarray}
R^-_\ell =\cases{R_{\ell 1},&$r<r_0$;\cr
                \alpha R_{\ell 3}+\beta R_{\ell 4},&$r>r_0$;\cr}\qquad
R^+_\ell =\cases{\gamma R_{\ell 1}+\delta R_{\ell 2},&$r<r_0$;\cr
                 R_{\ell 4},&$r>r_0$.\cr}
\end{eqnarray}
The coefficients are fixed by demanding that the functions and their first
derivatives be continuous at $r=r_0$; they may be expressed in terms of
Wronskians evaluated at $r=r_0$.

     When a particle moves under the influence of a highly nonspherical
potential, the Green function should be constructed in coordinates chosen
so it is reasonable for $F$ to depend only on one of them, $\xi$.  As an
example, consider a potential well in the shape of a right circular cylinder
of diameter $D$ and length $L$:
\begin{eqnarray}
V({\bf r})=\cases{-V_0&$|z|<L/2, \sqrt{x^2+y^2}<D/2$;\cr
               0&otherwise.\cr}
\end{eqnarray}
If $L\sim D$, this is an example of an approximately spherical potential,
and it is appropriate to use the Green function constructed above.
Here we consider the case of a long thin rod: $L>>D$.  Now it is manifestly
inadequate to take the surfaces of constant $F$ to be spherical.  These
surfaces ought to be figures of rotation about the $z$-axis in the form
of elongated cigars: prolate spheroids.  This suggests use of the prolate
spheroidal coordinate system:
\vfill
\eject
\begin{eqnarray}
x={r_0\over 2}\sqrt{\xi^2-1}\sin\theta\cos\phi, \nonumber \\
y={r_0\over 2}\sqrt{\xi^2-1}\sin\theta\sin\phi, \nonumber \\
z={r_0\over 2}\xi\cos\theta; \\
1\le\xi <\infty ,\qquad 0\le\theta\le\pi ,\qquad 0\le\phi <2\pi ; \nonumber \\
dV=\left ({r_0\over 2}\right )^3(\xi^2-\cos^2\theta )\sin\theta d\xi
d\theta d\phi.\nonumber
\end{eqnarray}
Surfaces of constant $\xi$ are ellipsoids of revolution about the z-axis, the
ellipses having foci at $z=\pm r_0/2$.  If $r_0$ is appropriately chosen (by
a variational minimization of $E_0$), the shape of the ellipsoids can be made
to resemble that of the rod for $\xi\sim 1$; for large $\xi$ the ellipsoids
approach spheres of radius $r_0\xi /2$.  An obvious choice for $F$ is
\begin{equation}
F(\xi )=-Kr_0\xi /2-\ln\xi.
\end{equation}

     In prolate spheroidal coordinates, Eq.~\ref{eq:Greenfun} takes the
form
\begin{eqnarray}
{{\partial}\over{\partial\xi}}\left [(\xi^2-1){{\partial G}\over{\partial\xi}}
\right ]+{1\over{\sin\theta}}{{\partial}\over{\partial\theta}}\left [\sin\theta
{{\partial g}\over{\partial\theta}}\right ]+\left [{1\over{\xi^2-1}}+{1\over
{\sin^2\theta}}\right ]{{\partial^2 G}\over{\partial\phi^2}} \nonumber \\
+2(\xi^2-1){{dF}\over{d\xi}}{{\partial G}\over{\partial\xi}}={2\over{r_0}}
\delta (\xi-\xi')\delta (\cos\theta -\cos\theta')\delta (\phi -\phi').
\end{eqnarray}
The solution has the form
\begin{equation}
G(\xi ,\theta , \phi ;\xi' ,\theta ',\phi ')=\sum_{\ell =0}^\infty\sum_{m= -
\ell}^\ell g_{\ell, m}(\xi ,\xi')Y_{\ell m}(\theta ,\phi )Y_{\ell m}^*(\theta'
\phi'),
\end{equation}
where the radial Green function satisfies
\begin{eqnarray}
(\xi^2-1){{\partial^2 g_{\ell m}}\over{\partial\xi^2}}+2\left [\xi +(\xi^2-1)
{{dF}\over{d\xi}}\right ]{{\partial g_{\ell m}}\over{\partial\xi}}-\left [
\ell (\ell +1)+{{m^2}\over{\xi^2-1}}\right ] g_{\ell m} \nonumber \\
={2\over{r_0}}\delta (\xi -\xi').
\label{eqn:xiequal}
\end{eqnarray}
The radial Green function is
\begin{equation}
g_{\ell m}(\xi,\xi' )={2\over{r_0C_{\ell m}}}e^{2F(\xi ')}\Xi_{\ell m}^-(\xi_<)
\Xi_{\ell m}^+(\xi_>).
\end{equation}
$C_{\ell m}$ appears in the Wronskian $W(\Xi^+,\Xi^-)=C_{\ell m}e^{-2F}/(\xi^2
-1)$.  The functions $\Xi_{\ell m}^\pm$ satisfy the homogeneous equation
obtained by removing the delta function source in Eq.~\ref{eqn:xiequal}.

     The general method of constructing the Green function appropriate to the
rod-shaped potential well is now clear.  We discontinue discussion of the
functions $\Xi_{\ell m}^\pm$ except to point out that the differential
equation they satisfy has three regular singular points and an irregular
singular point at infinity.

\section{Long Range Potentials}
\setcounter{equation}{0}

     When a potential is long range and behaves at large distance like
$r^{-p}$, $1<p<2$, there is no longer a threshold strength for the bound
state.\cite{landau3}  The anti-WKB approximation reproduces this result,
and it does so in lowest order, when $n=0$.  We demonstrate this
by showing that with the choices of $F$ given by Eq.~\ref{eqn:threeFs} we
can adjust $r_0$ so that $E_0$=0 no matter how weak the strength of the
potential tail.  A futher adjustment of $r_0$ will then produce a negative
(bound state) energy.

     We consider the case of $F_2$.  For the potential $V_0/r^p$, $(r>a)$, the
eigenvalue equation implied by Eq.~\ref{eqn:sphereigen} takes the form,
when $n=E_0=0$:
\begin{equation}
0=-{{2mV_0}\over\hbar^2}\int_a^\infty{{r^{2-p}dr}\over{r^2+r_0^2}}+3r_0^2
\int_0^{\infty}{{r^2 dr}\over{(r^2+r_0)^3}}.
\end{equation}
This may be written
\begin{equation}
0=-{{mV_0r_0^{1-p}}\over\hbar^2}{{2\pi}\over{2\sin\pi(p-1)/2}}-\sum_{k=0}
^\infty \left ({a\over r_0}\right )^{2k+3-p}{{(-1)^k}\over{k+(3-p)/2}}
+{{3\pi}\over{16r_0}}.
\end{equation}
For $1<p<2$, there is some large value of $r_0$ for which this equation is
satisfied, no matter how small $V_0$ may be.  There must be a nodeless bound
state.

     When $p=1$, we come to the classic case of the Coulomb potential.  At
this point, the asymptotic behavior of the wavefunction changes, and in
three dimensions the coefficient of the $\ln(r/r_0)$ term depends on the
potential strength and bound state energy.  Since this coefficient is no
longer known, the obvious response is to treat this coefficient as an
additional parameter, $r_1$.  Thus, for example, $F_3$ is generalized to
\begin{eqnarray}
F_3(r)=\cases{r_1-(K+{r_1\over{r_0}})r,&$(r<r_0);$\cr
               -Kr-r_1\ln (r/r_0),&$(r>r_0)$.\cr}
\end{eqnarray}
But now the determination of the variational parameters is obvious, because
with $r_1=0$, $F_3=-Kr=S.$  Therefore $T(r)=0$, and we immediately have the
hydrogen wavefunction, with $E_0=E$.  The same result is achieved with
$F_1$ and $F_2$.

     It is fortuitous that the hydrogen wavefunction occurs among the natural
choices for $F$.  Still, it is nice that the Coulomb potential is so easily
encompassed by the anti-WKB approximation.

\section{$\phi^4$ Field Theory}
\setcounter{equation}{0}

     The anti-WKB approximation can be applied to problems having many degrees
of freedom.  We illustrate this using the case of $\phi^4$ field theory
in one spatial dimension.  In the continuum the Hamiltonian is
\begin{eqnarray}
H=\int_0^L dx\Biggl \{ {1\over 2}c^2\pi^2(x)+{1\over 2}\left [{{d\phi (x)}\over
{dx}}\right ]^2 +{1\over 2}{c^2\over{\hbar^2}}(m^2+\delta m^2)\phi^2(x)
\nonumber \\
+{{\lambda}\over{24}}\phi^4(x)\Biggr \}.
\end{eqnarray}
We identify fields at $x=0$ and $x=L$.  To treat the system by the anti-WKB
method we must deal with a discrete set of degrees of freedom.  We therefore
divide the line into $N$ segments of length $a$; $N=L/a$.  Each segment is
represented by a dimensionless lattice coordinate $\phi_k$, $k=1,\dots ,N$.
The lattice Hamiltonian is
\begin{eqnarray}
H={{\hbar c}\over a}\sum_{k=1}^N\Biggl[-{1\over 2}{{\partial^2}\over{\partial
\phi_k^2}}+{1\over 2}(\phi_{k+1}-\phi_k)^2+{1\over 2}\left ({{ca}\over\hbar}
\right )^2(m^2+\delta m^2)\phi^2_k \nonumber \\
{{\lambda\hbar ca}\over{24}}\phi^4_k\Biggr ].
\label{eq:latham}
\end{eqnarray}

     Putting the field theory on the lattice introduces a short distance
cutoff $a$, or equivalently a large momentum cutoff $\hbar /a$.  This
removes the notorious divergences of continuum quantum field theory.  However,
the divergences still lurk and reveal themselves when we approach the
continuum limit by taking $a\to 0$, $N\to\infty$, with $Na=L$ fixed.  We
want to be able to take the limit, of course, since for us the lattice is only
a computational device.  In our case we find that as we take $a\to 0$ we
approach a continuum theory having infinite mass.  (We show this below.)  This
pathology disappears when we include an appropriate mass counterterm in the
Hamiltonian:
\begin{eqnarray}
\delta m^2=-{{\lambda\hbar}\over{4c}}\Delta_0;\qquad\Delta_0={1\over N}
\sum_{p=1}^N {1\over{\omega_0(p)}};\label{eq:counterterm} \\
\omega_0(p)=\sqrt{\left ({{mca}\over
\hbar}\right )^2+4\sin^2\left ({{\pi p}\over N}\right )}. \nonumber
\end{eqnarray}
This counterterm is the lattice version of the very counterterm that must be
included in the continuum Hamiltonian.  At small $a$, $\Delta_0$ grows
logaritmically to keep the effective mass finite.
\begin{equation}
\Delta_0\sim {1\over\pi}\ln {{\hbar^2}\over{m^2c^2a^2}}.
\label{eq:limit}
\end{equation}

     In one spatial dimension $\phi^4$ field theory is
superrenormalizable.~\cite{chang}  For this system, the explicit
counterterm in~\ref{eq:counterterm} suffices to remove all $a\to 0$
divergences in ``physical'' entities like the correlator
\begin{equation}
<0|\phi(0)\phi(x)|0>\sim\hbar c<0|{1\over N}\sum_{k=1}^N\phi_k\phi_{k+x/a}|0>.
\label{eq:correlator}
\end{equation}
Unfortunately, the groundstate energy $E$ that has figured prominently in the
anti-WKB approximation is not ``physical'' in the sense used here.  Even
free field theory, $\lambda =0$, includes zero point energies for each
degree of freedom, and when summed these diverge as $a\to 0$.  We will cope
with $E$ as we go along.  This point aside, Eqs.~\ref{eq:latham},~\ref
{eq:counterterm} define a lattice representation of continuum $\phi^4$ field
theory to which we can apply the anti-WKB approximation.

     A central issue for this many degree of freedom problem is finding an $F$
that is appropriate, and for which the multidimensional integrals and Green
function can be found.  There is just one choice for which this is {\it easy}:
\begin{equation}
F=-{1\over 2}\sum_{k_1k_2}M_{k_1k_2}\phi_{k_1}\phi_{k_2}\equiv -
{1\over 2}\phi M\phi,
\end{equation}
with $M$ a real symmetric matrix.  Consider the eigenvalue equation for $E_0$.
\begin{eqnarray}
0=\int(\Pi d\phi )e^{-\phi M\phi}\Bigl\{\sum_k\Bigl [(\phi_{k+1}-\phi_k)^2
+(ca/\hbar)^2(m^2+\delta m^2)\phi_k^2 \nonumber \\
+(\lambda\hbar ca^2)\phi_k^4/12\Bigr ]-2E_0a/\hbar c+Tr(M)-\phi M^2\phi\Bigr\}.
\label{eq:E0}
\end{eqnarray}
Every term in the brace involves powers of $\phi_k$, so the integral can be
evaluated by taking derivatives with respect of $\alpha_k$ of the generating
function
\begin{equation}
I_1(\alpha )=\int(\Pi d\phi)\exp (-\phi M\phi+i\alpha\phi );\qquad \alpha\phi
\equiv \sum_k\alpha_k\phi_k.
\end{equation}
This generator, in turn, can be evaluated easily by a change of coordinates.
Let the eigenvectors and eigenvalues of $M$ be $v^q$ and $\mu^q$.
\begin{equation}
Mv^q=\mu^qv^q,\qquad (q=1,\dots N).
\end{equation}
The new coordinates are
\begin{equation}
\eta^q=\sum_k\phi_kv_k^q;\qquad \phi_k=\sum_q\eta^qv_k^q.
\end{equation}
In these coordinates $I_1$ becomes a product of $N$ independent Gaussian
integrals, and
\begin{equation}
I_1(\alpha )={{\pi^{N/2}}\over{\sqrt{detM}}}\exp\left (-{1\over 4}\alpha M^{-1}
\alpha\right ).
\end{equation}

     Eq.~\ref{eq:E0} becomes
\begin{eqnarray}
0=I_1(0)\Bigl\{\sum_k\Bigl [M^{-1}_{kk}-M^{-1}_{k,k+1}+(ca/\hbar)^2
(m^2+\delta m^2)M^{-1}_{kk}/2 \nonumber \\
+(\lambda\hbar ca^2)(M^{-1}_{kk})^2/16\Bigr ]-2E_0a/\hbar c+Tr(M)-
Tr(M^2M^{-1})\Bigr\}.
\end{eqnarray}
$M_{k_1k_2}$ must be a function of $k_1 -k_2$ of period $N$ owing to the
homogeneity and periodicity of the Hamiltonian.  It can be written
as a discrete Fourier transform.
\begin{equation}
M^{\pm 1}_{k_1k_2}={1\over N}\sum_{p=1}^N\omega^{\pm 1} (p)\exp \left [{{2\pi
ip(k_1-k_2)}\over{N}}\right ].
\label{eq:planewaves}
\end{equation}
With this the eigenvalue equation decouples into modes.
\begin{eqnarray}
0=I_1(0)\Biggl\{\sum_p\Biggl [{{\omega (p)}\over 2}+{1\over{\omega (p)}}\Biggl
 (1-\cos {{2\pi p}\over N}+(ca/\hbar)^2(m^2+\delta m^2\Biggr)\Biggr ]
\nonumber \\ +N(\lambda\hbar ca^2)\Biggl [{1\over N}\sum_p
{1\over{\omega (p)}}\Biggr ]^2-{{2E_0a}\over{\hbar c}}\Biggr\}.
\end{eqnarray}

     The $\omega (p)$ are parameters in $M$ that are fixed by minimizing
$E_0$.  We find
\begin{eqnarray}
\omega (p)=\sqrt{\left ({{ca}\over \hbar}\right )^2(m^2+\delta m^2)+
+{{\lambda\hbar ca^2}\over 4}\Delta +\sin^2\left ({{\pi p}\over N}\right )}, \\
\Delta={1\over N}\sum_p {1\over{\omega (p)}}.
\label{eq:delta}
\end{eqnarray}

     Eq.~\ref{eq:delta} determines $\Delta$, and it is enlightening to examine
its solution near the continuum limit {\it when we drop the counterterm}
$\delta m^2$.  Following Eq.~\ref{eq:limit},
\begin{equation}
\Delta\sim {1\over\pi}\ln{4\over{\lambda\hbar ca^2\Delta}};\qquad
\Delta\sim {1\over\pi}\ln{4\over{\lambda\hbar ca^2}}+O\left (\ln\ln{4\over
{\lambda\hbar ca^2}}\right ).
\end{equation}
Therefore, the square of the effective mass in $\omega (p)$ becomes, in the
limit of small $a$,
\begin{equation}
m^2+{{\lambda\hbar^3\Delta}\over{4c}}\sim m^2+{{\lambda\hbar^3}\over{4\pi c}}
\ln {4\over{\lambda\hbar ca^2}}.
\end{equation}
This shows that we approach a continuum theory having infinite mass.  On the
other hand, when we do include the counterterm, Eq.~\ref{eq:delta} has the
solution $\Delta=\Delta_0$; $\omega (p)=\omega_0 (p)$.  Then the
mode function $\omega (p)$ never changes, no matter what the lattice spacing
or coupling strength.

     The energy $E_0$ is
\begin{equation}
{E_0\over L}={{\hbar c}\over{2a^2}}\left [{1\over N}\sum_{p=1}^N\omega(p)
\right ]-{{m^2c^3}\over{\hbar}}\left ({{\lambda\hbar^3}\over{m^2c}}\right )
\Delta_0^2.
\end{equation}
The sum on the right is the zero-point energy of the degrees of freedom.  It
is present in free field theory, and its contribution is quadratically
divergent in $a^{-1}$.  The last term, proportional to the dimensionless
coupling $\lambda\hbar^3/m^2c$, diverges like $(\ln a^{-1})^2$.  In
perturbation theory it arises because anomalous combinatorics spoil the
cancellation of divergences in self-energy loops and $\Delta_0$.  This
failure to cancel occurs only at order $\lambda$ of perturbation theory, so
$E_0$ already exhibits all the terms that diverge as $a\to 0$.

     To compute $E_1$ we need the Green function satisfying
\begin{equation}
\left ({{\partial^2}\over{\partial\phi^2}}-2\phi M{{\partial}\over{\partial
\phi}}\right )G(\phi ,\phi' )=\prod_k\delta (\phi_k-\phi_k'),
\end{equation}
in matrix notation.  We use Eq.~\ref{eq:Greenexp} to construct $G$, which
requires us to construct eigenfunctions of the operator on the left.  We
again use coordinates $\eta$; then the equation for the eigenfunctions and
eigenvalue is
\begin{equation}
\sum_{q=1}^N\left [{{\partial^2}\over{\partial (\eta^q)^2}}-2\mu^q\eta^q
{{\partial}\over{\partial\eta^q}}\right ]\psi=\mu\psi.
\end{equation}
This equation may be solved by separation of variables.
\begin{equation}
\psi_{\{s\}}=\prod_{q=1}^N H_{s_q}(\eta^q\sqrt{\mu^q}).
\end{equation}
The functions $H_s$ are Hermite polynomials satisfying
\begin{eqnarray}
{{d^2H_s(x)}\over{dx^2}}-2x{{dH_s(x)}\over{dx}}+2sH_s(x)=0, \\
\int_{-\infty}^{\infty}dxe^{-x^2}H_{s_1}(x)H_{s_2}(x)=\delta_{s_1,s_2}
\sqrt{\pi}2^{s_1}(s_1!). \nonumber
\end{eqnarray}
There are $N$ indices on our wave function, and the eigenvalue is a linear
function of them:
\begin{equation}
\mu=-2\sum_{q=1}^Ns_q\mu^q.
\end{equation}
The Green function is
\begin{eqnarray}
G=\left (-{1\over 2}\right )e^{-\phi'M\phi'}\int_0^1{{dz}\over z}\prod_{q=1}^N
\sqrt{{{\mu_q}\over\pi}}\Biggl[\sum_{s_q=0}^{\infty}\left ({{z^{\mu_q}}\over 2}
\right )^{s_q} \\
\times{{H_{s_q} (\eta^q\sqrt{\mu^q})H_{s_q} ({\eta^q}'\sqrt{\mu^q})}
\over{(s_q!)}}\Biggr ]. \nonumber
\end{eqnarray}
The purpose of the integration over $z$ is to produce the denominator
$1/\mu_k$ in Eq.~\ref{eq:Greenexp}.  The sum over $s_q$ is given by Mehler's
formula.~\cite{erdleyi}  Reverting to the original variables,
\begin{eqnarray}
G(\phi ,\phi' )=-{1\over 2}{{\sqrt{det(M)}}\over{\pi^{N/2}}}\int_{\epsilon}^1
{{dz}\over{z\sqrt{det(1-z^{2M})}}}
\exp\Biggl[-\phi'{M\over{1-z^{2M}}}\phi' \nonumber \\
-\phi{{Mz^{2M}}\over{1-z^{2M}}}\phi
+2\phi'{{Mz^M}\over{1-z^{2M}}}\phi\Biggr].
\label{eq:A}
\end{eqnarray}

     As we expect from the discussion of Section 2, this Green function does
not exist because $\mu=0$ is an eigenvalue.  In Eq.~\ref{eq:A} the divergence
appears at $z=0$, and it has been regulated by a cutoff at $z=\epsilon$.
However, when computing $T_n$, the coefficient of the $1/z$ factor is
\begin{eqnarray}
\int(\prod d\phi )e^{-\phi M\phi}D\phi ; \\
D=\Biggl\{ \sum_k\Biggl [(\phi_{k+1}-\phi_k)^2+\left ({{ca}\over{\hbar}}
\right )^2(m^2+\delta m^2)\phi_k^2 \nonumber \\
+{{\lambda\hbar ca^2}\over{12}}\phi_k^4\Biggr ]-{{2aE_n}\over{\hbar c}}+Tr(M)
-\phi M^2\phi-
\left ({{\partial T_{n-1}}\over{\partial\phi}}\right )^2\Biggr\}. \nonumber
\end{eqnarray}
Setting this to zero, we obtain a finite $T_n$ as $\epsilon\to 0$.  Thus we
again encounter the eigenvalue equation as a consistency condition for the
solution of Schr\"odinger's equation.

     The dependencies on the fields in Eq.~\ref{eq:A} are Gaussian, so we can
use $I_1$ to evaluate the generating function
\begin{eqnarray}
I_2(\alpha ,\beta )=\int(\prod d\phi d\phi')G(\phi ,\phi')\exp (-\phi M\phi+i
\alpha\phi+i\beta\phi' ) \nonumber \\
={{\pi^{N/2}}\over{\sqrt{det(M)}}}\exp\left (-{1\over 4}\alpha M^{-1}\alpha-
{1\over 4}\beta M^{-1}\beta\right ) \\
\times \left (-{1\over 2}\right )\int^1_{\epsilon}{{dz}\over z}\exp \left (
-{1\over 2}\alpha[z^MM^{-1}]\beta\right ). \nonumber
\end{eqnarray}
By differentiation of $I_2$ we obtain all the contributions in
\begin{equation}
E_1-E_0={{\hbar c}\over{2a}}{{\int (\Pi d\phi d\phi')e^{-\phi M\phi}G(\phi,
\phi')D(\phi )D(\phi' )}\over{\int(\Pi d\phi )e^{-\phi M\phi}}}.
\end{equation}
The result is
\begin{equation}
{{E_1-E_0}\over L}=-{{m^2c^3}\over{\hbar}}\left ({{\lambda\hbar^3}\over{m^2c}}
\right )^2{{\Delta_2}\over{384}},
\end{equation}
where
\begin{equation}
\Delta_2=\left ({{mca}\over{\hbar}}\right )^2\int_0^1{{dz}\over z}{1\over N}
\left [\sum_{k_1,k_2}\left (z^MM^{-1}\right )_{k_1,k_2}\right ]^4.
\end{equation}
Using Eq.~\ref{eq:planewaves},
\begin{equation}
(z^MM^{-1})_{k_1,k_2}={1\over n}\sum_{p=1}^N{{z^{\omega_0(p)}}\over{\omega_0
(p)}}\exp\left [{{2\pi ip(k_1-k_2)}\over N}\right ],
\end{equation}
\begin{eqnarray}
\Delta_2=\left ({{mca}\over\hbar}\right )^2\left ({1\over N}\right )^3
\sum_{p_1,p_2,p_3}{1\over{\omega_0(p_1)\omega_0(p_2)\omega_0(p_3)
\omega_0(p_1+p_2+p_3)}} \nonumber \\
\times{1\over{[\omega_0(p_1)+\omega_0(p_2)+\omega_0(p_3)
+\omega_0(p_1+p_2+p_3)]}}.
\end{eqnarray}
It is straightforward to verify that $\Delta_2$ remains finite as $a\to 0$.

     What emerges after all our work appears to be perturbation theory in
``old-fashioned'' form.  This is easily demonstrated.  We have $E_1-E_0=
O(\lambda^2)=O[(\partial T_0/\partial\phi )^2]$.  Hence, $T_0=O(\lambda )$.
Then use
\begin{eqnarray}
T_n-T_{n-1}=\int(\Pi d\phi' )G(\phi,\phi ')\Biggl [{{2a}\over{\hbar c}}(E_{n-1}
-E_n) \nonumber \\
+\left ({{\partial T_{n-2}}\over{\partial\phi}}\right )^2
-\left ({{\partial T_{n-1}}\over{\partial\phi}}\right )^2\Biggr ], \\
E_n-E_{n-1}={{\int (\Pi d\phi )e^{2F}[(\partial T_{n-2}/\partial\phi )^2
-(\partial T_{n-1}/\partial\phi )^2]}\over{\int (\Pi d\phi )e^{2F}}}
\nonumber
\end{eqnarray}
to prove by induction $T_n-T_{n-1}=O(\lambda^{n+1})=E_n-E_{n-1}$.  This
result is not surprising.  Our $F$ is the $S$ of free field theory, and
we compute corrections to that.  This is the program of perturbations theory.

     No one wants to use the anti-WKB approximation to develop
perturbation theory.  What we learn from our exercise is that the calculations
can be carried out, and they lead to a sensible and familiar result.  But it
is possible, within the anti-WKB approximation,
to contemplate alternatives that transcend perturbation theory.  An obvious
generalization of $F$ is
\begin{equation}
\tilde F=\sum_{q=1}^N\left [-{1\over 2}\tilde\mu^q(\eta^q)^2-{1\over 4}
\tilde\nu^q(\eta^q)^4\right ].
\end{equation}
We use plane wave coordinates so that all degrees of freedom are coupled.
The mode eigenvalue equation is unfamiliar, and Mehler's formula is
not available, so we must work harder.  A computer would be required to
assemble $G$.

     A payoff is possible when a seed like $\tilde F$ is used.  Recall that at
$\lambda\hbar^3/m^2 c\sim 10$, $\phi^4$ field theory makes a transition to
a state of broken $\phi \leftrightarrow -\phi$ symmetry in which
the field has a vacuum expectation value:
$<0|\phi (x)|0>=\phi_0\ne 0$.~\cite{chang}  From this fact alone we see that
$F$ must be inadequate at strong coupling because with $F$, $\phi_0=0$.  But
in $\tilde F$, as the variational parameters $\tilde\mu^q$ vary smoothly with
$\lambda$, a vacuum expectation value can develop beyond a critical coupling.

     A trick attibuted to Feynman applies to this problem.  Extend H by adding
a term $\xi(1/N)\sum_k\phi_k\phi_{k+d}$.  The extended Hamiltonian remains
homogeneous and periodic, so the techniques we have developed continue to
apply.  $E$ depends on $\xi$, of course, and by first order perturbation
theory (in $\xi$), $dE/d\xi(\xi =0)$ is just the correlator of
Eq.~\ref{eq:correlator}.  In this formula we would substitute $E_n$ for $E$.

\section{Helium}
\setcounter{equation}{0}

     Because of the Pauli principle, helium is the only multi-electron atom in
which a symmetric spatial wave function is physical.  The nodeless state we
construct is the groundstate of helium.

     The Hamiltonian we study includes only the potential terms representing
the electric forces acting on electrons moving around an immobile nucleus.
The ground state energy of the simplified Hamiltonian therefore differs
slightly from that of physical helium where magnetic, recoil and relativistic
corrections are present.  The Hamiltonian is
\begin{equation}
H=-{1\over 2}(\nabla_1^2+\nabla_2^2)-{2\over r_1}-{2\over r_2}+
{1\over{|{\bf r}_1-{\bf r}_2|}}.
\end{equation}
We use dimensionless coordinates, so the eigenvalues of $H$ must be multiplied
by $me^4/\hbar^2$.

     A standard textbook variational computation with this Hamiltonian
assumes that the wavefunction is a product of exponential (hydrogenic)
factors.~\cite{landau4}  The resulting energy, $<H>=-(27/16)^2=-2.848$, is
very close to the experimental groundstate energy, -2.90, and we adopt it as
a standard for the problem.  Since we have the the tools to deal with
many-body problems when $F$ is a quadratic form, we use that:
\begin{equation}
F=-{1\over 2}K_0(r_1^2+r_2^2)-K_1{\bf r}_1{\bf\cdot r}_2.
\label{eq:form}
\end{equation}
Our choice of $F$ corresponds to a Gaussian first guess for
the helium wavefunction.  We compute $E_0$, choosing $K_0$ and $K_1$ to
minimize it and find
\begin{equation}
E_0=-2.324;\qquad K_0=1.549\qquad K_1=(0.09238)K_0.
\end{equation}
We obtain only 0.816 of the ``standard'' binding, which means that the
anti-WKB correction is substantial.

     An interesting feature of our result is that $K_1$ is so small.  The
repulsion between electrons enhances the probability of finding
the electrons on opposite sides of the nucleus, but not by much.  In fact, if
we set $K_1=0$, the binding is decreased by less than 1\%, and $E_0=-2.301$.
We use this version of $F$ when calculating $E_1$ because of the resulting
simplifications.  The computation of $E_1$ now parallels that of Section 5.
The main new feature is that the terms in the potential are Coulomb, not
polynomial.  When we use generating function $I_2$, we integrate over
parameters to generate the Coulomb terms, using the relation
\begin{equation}
{1\over r}={1\over {2\pi^2}}\int{{d^3\alpha}\over{\alpha^2}}
\exp [i{\bf\alpha\cdot r}].
\end{equation}
Summarizing our results:
\begin{eqnarray}
E_0=-2.301;\quad E_1=-2.707,\quad <H>=-2.848; \nonumber \\
{{E_0}\over {<H>}}=0.808,\qquad {{E_1}\over {<H>}}=0.951.
\end{eqnarray}

\section{Conclusions}
\setcounter{equation}{0}

     We have proposed an approximation for quantum mechanics and
field theory that assumes the relative importance of gradient and Laplacian
terms is the reverse of that in the WKB approximation.  As with any such
scheme, usefulness is an important issue.  We have explored usefulness by
applying the anti-WKB approximation to several problems.  The examples of the
spherical square well and the helium atom provide interesting numerical
results.  In these cases we chose a simple variational seed $F$
and found that the initial (variational) estimates of the ground state
energy were incorrect by several tens of percent.  The first nontrivial
anti-WKB correction removed three quarters of the error, or more.

     In practice, the WKB method is difficult to apply to problems with many
degrees of freedom.  The anti-WKB method does suffer from this fault.  With a
quadratic choice for $F$ we developed the formulas necessary for the treatment
of the rather different problems of $\phi^4$ field theory and the helium
atom.

\bigskip

\noindent{\large{\bf Acknowledgements}}

This work was supported in part by the National Science Foundation under
grant number NSF-PHY-91-21039.

\bibliographystyle{plain}
\vfill
\eject

\end{document}